# Silicon Carbide Photonic Crystal Cavities with Integrated Color Centers

Greg Calusine[1,a)], Alberto Politi[1,a),b)], and David D. Awschalom[1,2]

1. Department of Physics, University of California, Santa Barbara, CA 93106, USA
2. Institute for Molecular Engineering, University of Chicago, Chicago, IL 60637, USA

Abstract

The recent discovery of color centers with optically addressable spin states in 3C silicon carbide (SiC) similar to the negatively charged nitrogen vacancy center in diamond has the potential to enable the integration of defect qubits into established wafer scale device architectures for quantum information and sensing applications. Here we demonstrate the design, fabrication, and characterization of photonic crystal cavities in 3C SiC films with incorporated ensembles of color centers and quality factor (Q) to mode volume ratios similar to those achieved in diamond. Simulations show that optimized H1 and L3 structures exhibit Q's as high as 45,000 and mode volumes of approximately $(\lambda/n)^3$. We utilize the internal color centers as a source of broadband excitation to characterize fabricated structures with resonances tuned to the color center zero phonon line and observe Q's in the range of 900-1,500 with narrowband photoluminescence collection enhanced by up to a factor of 10. By comparing the Q factors observed for different geometries with finite-difference time-domain simulations, we find evidence that nonvertical sidewalls are likely the dominant source of discrepancies between our simulated and measured Q factors. These results indicate that defect qubits in 3C SiC thin films show clear promise as a simple, scalable platform for interfacing defect qubits with photonic, optoelectronic, and optomechanical devices.

Silicon carbide (SiC) has recently emerged as a promising material for the integration of defect qubit states into microfabricated and nanofabricated devices. The three most prevalent crystalline forms of SiC-termed 4H, 6H, and 3C-have all demonstrated deep-level defect states (color centers) with spin and optical properties similar to the negatively charged nitrogen vacancy center in diamond but with luminescence bands in the telecom wavelength range.[1,2] As a potential host material for defect qubits, all three crystal forms share many of diamond's favorable properties such as a wide band gap, the potential for isotopic purification, and high thermal conductivity.[3,4] Moreover, the various forms of SiC also have advantages for the potential incorporation of defect qubit states into conventional semiconductor devices. Material

---

growth and microfabrication techniques are much more mature in SiC, and the material can be controllably doped to be either n- or p-type.[5]

The cubic form of SiC, 3C, has the additional advantage of being commercially available as a high quality heteroepitaxial thin film grown on wafers of silicon. This provides for a convenient route towards improving the measurement and control of color center defect qubits by altering their local photonic environment because they are typically initialized and measured through the use of their spin dependent optical dynamics. Accordingly, this allows for the incorporation of defect qubits in 3C SiC into a wide variety of optical devices for use in developing on-chip single photon sources,[6,7] integrated lasers,[8] or opto-mechanical devices.[9] Many of these applications require or are facilitated by the use of high quality factor ('Q'), small mode volume cavities to enhance light-matter interactions for optical transitions within the color center zero phonon line.[10] In particular, microcavities can enhance single photon emission into a single, well-defined mode,[11] increase the rate of photon emission,[12] or achieve coherent interactions between optical transitions and cavity modes in the strong coupling regime.[13] While much is still unknown regarding the optical and electronic properties of defect qubit states in 3C SiC, the possibility of integrating defects with such a broad range of functional devices makes this material a promising platform for quantum information and sensing applications.[14]

In this work, we present the design, fabrication, and characterization of small mode volume photonic resonators[15,16] based on 3C SiC heteroepitaxial films on silicon operating throughout the emission band of the Ky5 color center (approximately 1100 to 1300 nm)[17] that exhibit narrowband collection enhancements as high as a factor of ten. The Ky5 color center is one potential defect qubit candidate in 3C SiC with optical and electronic properties similar to the negatively charged nitrogen vacancy center in diamond and has been previously suggested to be a form of divacancy.[18] In order to produce resonant modes to couple to Ky5 center luminescence, we design structures consisting of one and three missing holes in a triangular array of holes in a freestanding dielectric slab, commonly referred as H1 and L3 and depicted in Figure 1(a) and Figure 1(c), respectively. Finite-difference time-domain (FDTD) simulations (MEEP,[19] Lumerical) were used to determine the structure parameters necessary to produce modes within the spectral range of the zero phonon line of the Ky5 optical transitions and to optimize the position of the holes around the cavity to improve the cavity Q. For both the H1 and L3 cavity types, the lattice hole radius 'r' and lattice periodicity 'a' are fixed by the relation r=0.29a. Following the method in reference 20, we obtained an optimized Q of ~ 45,000 for the H1 structure when the radius of the four nearest vertically adjacent holes to the cavity are reduced to 0.22a and the horizontally adjacent holes are displaced outwards by a distance 0.2a. The optimal Q is obtained with a film thickness of 0.8a. Simulations indicate that the theoretical mode volume is ~ $(\lambda/n)^3$. We similarly optimized the L3 structure by shifting the position of the side holes of the cavity[21] by 0.21a and reducing their size[22] by 0.12a. This results in a theoretical quality factor Q ~ 17,000 at a film thickness of 0.85a with mode volume ~ 0.9 $(\lambda/n)^3$. These values compare favorably to the unoptimized L3 cavity design which achieves a Q factor of ~ 1,600 for a mode volume of $(\lambda/n)^3$.

The starting material for photonic crystal cavity fabrication consists of a 1 micron thick film of commercially available, <100> oriented 3C silicon carbide grown epitaxially on a 100 mm <100> oriented silicon wafer [NovaSic, see reference 23]. The initial film is thinned down to a thickness of 300 nm through a series of $SF_6$ and Ar/Cl inductively coupled plasma (ICP) etches. In order to produce a vacancy distribution vertically centered in the middle of the film, the film is then implanted with $^{12}C$ ions at an energy of 110 keV, an incidence angle of 7 degrees, and a dose of $1x10^{13}$ ions per square centimeter. The film is then annealed at 750°C for 30 minutes in an argon atmosphere to induce vacancy migration in order to promote the formation of Ky5 color centers. This implantation and annealing procedure was empirically determined to be optimal for producing a high density of Ky5 centers in the initially color center-free material.

In order to fabricate the photonic crystal structure, the 300 nm SiC film is first covered with a hard mask consisting of 100 nm of aluminum capped with 10 nm of titanium that is deposited onto the sample using electron beam evaporation. The titanium layer prevents oxidization of the aluminum and improves process consistency. The sample is then spin-coated with 340 nm of ZEP520 electron beam lithography resist and a conductive polymer layer (AQUASave) to eliminate charging effects during the resist exposure. Multiple arrays of devices are patterned on a 5-by-5-mm sample using a 100 keV electron beam lithography system. The conductive polymer is removed prior to the development of the electron beam resist and the pattern is then transferred to the SiC layer through a multistep ICP etch. First, a $BCl_3/Cl_2$ etch transfers the developed resist pattern to the Al/Ti layer. Then, without removing the sample from the etch chamber, an $SF_6$ plasma etch is performed at an ICP power of 900 W and a bias of 200 W to transfer the pattern to the SiC. This etch process was found to minimize the degradation of the hard mask edges during the aggressive but nearly vertical $SF_6$ etching of the silicon carbide layer.[24] The hard mask layers and any etch by-products remaining on the hole sidewalls are then removed through subsequent wet etches in Ti etchant, Al etchant, and buffered hydrofluoric acid. Finally, the sample is exposed to a short gaseous $XeF_2$ etch to isotropically remove the silicon substrate directly underneath the photonic crystal pattern to produce a freestanding structure. If necessary, the cavity resonance can then be fine-tuned by reducing the structure thickness with short $SF_6$ or Ar/Cl ICP etch steps. A complete process flow of the fabrication procedure is shown in Figure 2.

Scanning electron micrographs of the fabricated structures are shown in Figure 1(b) and Figure 1(d) for the optimized H1 and L3 designs, respectively. The structures exhibit a sidewall angle of approximately 85 degrees, typical of the values observed in the literature for similar etch conditions.[25] The periodic holes defining the structure are uniform in radius and accurate in their positioning to within the spatial resolution of the SEM (~ 1.5 nm). The thin film's surface roughness increases from 0.5 nm rms to 1.2 nm rms as a result of processing, as determined by atomic force microscopy. Small variations in the hole shapes with an rms radial deviation of 2.75 nm are typically observed and the hole sidewalls show slight striations likely due to degradation of the aluminum hard mask edges during the high power etch needed to produce vertical sidewalls. For each photonic cavity design, we fabricated an array of devices with varying parameters (lattice constant and hole radius) to geometrically tune the cavity mode over the desired wavelength range and determine the optimal structural parameters for achieving the highest Q.

The photonic crystal cavities were characterized in a home built scanning confocal microscope equipped with a helium flow cryostat. A 1060-nm diode laser was used for off-resonant excitation of color center luminescence through the Ky5 defects' blue-shifted absorption sideband and a custom Littman-Metcalf tunable diode laser was used for cross-polarized resonant scattering spectroscopy on the structures.[26] The incorporation of color centers directly into the cavity allows us to use the off-resonantly pumped defect luminescence band to excite the cavity modes over a broad range of wavelengths that are inaccessible to commercially available superluminescent diodes. Cross-polarized resonant scattering provides a complementary characterization method for probing cavities without incorporated emitters or to study the response of the cavity[11] or internal emitters[27] to an externally incident Gaussian input beam. Through the use of a fast-steering mirror incorporated into the optical path, both methods can provide a spatial map of the optical response of the photonic structure with ~ 1 micron resolution (Figure 1(e) and (f)) and for position feedback stabilization. Photoluminescence and scattered laser light were collected through the objective used for excitation and passed to a spectrometer or a series of sensitive photodetectors through a polarizing beam splitter.[23]

Figure 3(a) shows a typical broadband photoluminescence spectrum (red line) at low temperature (20K) resulting from off-resonant excitation of color centers in an unprocessed 1 micron thick 3C silicon carbide layer that has been implanted and annealed. The spectrum consists of a zero phonon line centered at 1118 nm and a broad phonon-assisted emission sideband that extends deeper into the infrared. Also shown in the figure is a spectrum (black line) originating from a 300 nm thick H1 photonic crystal cavity with incorporated Ky5 defects that has a cavity resonance around 1180 nm. While qualitatively similar, the zero phonon line broadens from 4.6 nm FWHM to 28.2 nm FWHM, likely due to increased inhomogeneous broadening caused by the decreased crystalline quality of the material grown closer to the silicon interface.[28]

Figure 3(b) shows the cavity resonance of an unoptimized L3 cavity with the resonance wavelength tuned to the peak of the Ky5 zero phonon line. The measured Q of 900 is less than the value of ~ 1,600 predicted by FDTD simulations. In the same plot, we show the cross-polarized resonant scattering spectrum of the same cavity, confirming the cavity properties measured via photoluminescence. Optimizations of the L3 cavity improve the measured cavity Q over the unoptimized case, as shown in Figure 3(c), but the resulting value is approximately an order of magnitude smaller than the predicted Q of ~ 17,000. Similarly, for the optimized H1 cavity design we observed a maximum Q of ~ 1,000 despite a simulated value of ~ 45,000 (Figure 3(d)). This design does, however, provide a greater degree of far field coupling than the other designs, resulting in a 10-times enhancement of narrowband photoluminescence collection as compared to emission falling outside the spectral range of the cavity. Additionally, we observed that the H1 cavity supports additional modes red-shifted from the fundamental with similar Q's but much smaller mode volumes.[23] In the case of the fundamental mode in the optimized H1 cavity, the cross-polarized resonant scattering spectrum exhibits a far more asymmetric Fano lineshape, indicating a greater degree of coupling between the cavity-dependent and -independent scattering channels. This is likely due to suppressed far field coupling to external Gaussian modes as a result of the odd cavity mode parity along the y-axis (as defined in Figure 1).

A number of factors may contribute to the observed discrepancy between the simulated and measured Q's of our fabricated structures. Direct absorption by the defects' optical transitions is unlikely to be a primary limitation of the experimentally measured Q's because the structures' optical resonances show similar Q's at room temperature, in films with lower defect densities, and in films that were not irradiated and annealed to form Ky5 emitters. Furthermore, based on the estimated absorption coefficient in material similar to ours (as measured in reference 29), our reduction in Q is likely not dominated by other material absorption processes. Instead, various structure imperfections inherent to the fabrication process-such as sidewall angle, variations in the size or position of the holes of the photonic crystal, or roughness on the surfaces or sidewalls-are the most likely source of scattering losses in excess of the radiative losses intrinsic to the cavity designs.

Table I shows the simulated Q for a given cavity geometry with absorption or geometric imperfections added to the simulated structure. Adding small, normally distributed random variations in the photonic crystal hole radii and positions to the simulations results in only small deviations from the ideal structure Q factor for the L3 cavities (1%-5%) and slightly larger deviations for the H1 design (up to 26%). The magnitude of these variations was adjusted to match the values observed from image analysis of SEM micrographs that corresponded to the resolution limit of the SEM of 1.5 nm and should be considered an upper bound on the actual values for the fabricated devices. The modest predicted reduction in Q's makes it unlikely that this contribution is dominant. Furthermore, the surface and sidewall roughness of 1.2 nm and 2.75 nm rms, respectively, that we observe in our fabricated devices is not likely to limit cavities with Q's of less than $10^5$.[30,31] Sidewall angle, however, has a pronounced effect on the Q factor for all three structures. Given the sidewall angle of approximately 85 degrees that we observe with SEM imaging, it is very likely that this is the limiting factor in reducing our Q's from the expected values.[32] Simulations predict that for sidewall angles less than 85 degrees, the optimized H1 cavity has a lower Q than that of the optimized L3, contrary to what is predicted for vertical sidewalls.[23] This qualitatively matches the behavior we observe in our fabricated structures.

In conclusion, we have demonstrated the design, fabrication and characterization of small mode volume photonic crystal microcavities in 3C silicon carbide thin films with integrated color center ensembles. We demonstrate modes with Q's of up to 1,500 and photoluminescence collection efficiency enhancements of up to a factor of 10. Comparisons of the simulated ideal cavity Q's with experiment and those simulated with fabrication imperfections suggest that the nonvertical sidewall angle of the hole etch is likely the dominant factor limiting the Q's of our fabricated structures. In order to produce high-Q photonic crystal cavities suitable for significant Purcell enhancements[12] or to reach the onset of strong coupling[13] to single defects, further work is needed to improve the verticality of the etch process step[33] or develop cavity designs that are less sensitive to sidewall angle. Alternatively, cavities with modest Q's can be used to achieve locally enhanced light-matter interactions[34] or for non-linear optics applications.[35] In particular, the second-order nonlinear susceptibility of SiC[36,37] enables the possibility of on-chip use of non-linear optical processes for frequency conversion or to overcome spectral inhomogeneity.[38] The ability to easily incorporate defect qubits into photonic, opto-electronic, and optomechanical devices on the wafer scale makes 3C silicon carbide a promising on-chip platform for future applications in the field of quantum information.

This work was supported by the AFOSR and NSF. A portion of this work was done in the UC Santa Barbara nano-fabrication facility, part of the NSF funded NNIN network. We acknowledge support from the Center for Scientific Computing at the CNSI and MRL.

## References

1. W. F. Koehl, B. B. Buckley, F. J. Heremans, G. Calusine & D. D. Awschalom, *Nature* 479, 84 (2011).
2. A. L. Falk, B. B. Buckley, G. Calusine, W. F. Koehl, V. V. Dobrovitski, A. Politi, C. A. Zorman, P. X.-L. Feng, and D. D. Awschalom, *Nat Commun* 4, 1819 (2013).
3. G.L. Harris (Ed.), *Properties of Silicon Carbide*, INSPEC, London (1995).
4. J.R Weber *et al.* Quantum computing with defects. *Proc. Natl. Acad. Sci. USA* **107**, 8513 (2010).
5. C.M. Zetterling (ed.) *Process Technology for Silicon Carbide Devices* (Institution of Electrical Engineers, 2002).
6. F. Fuchs, V. A. Soltamov, S. Väth, P. G. Baranov, E. N. Mokhov, G. V. Astakhov and V. Dyakonov, *Sci. Rep.*, 2013, 3, 1637.
7. S. Castelletto, B. C. Johnson, V. Ivády, N. Stavrias, T. Umeda, A. Gali and T. Ohshima, *Nature Mater.* 13, 151-156 (2014).
8. O. Painter, R. K. Lee, A. Scherer, A. Yariv, J. D. O'Brien, P. D. Dapkus, and I. Kim, *Science* 284(5421), 1819–1821 (1999).
9. P. Ovartchaiyapong, K. W. Lee, B. A. Myers, A. C. Bleszynski Jayich, arXiv:1403.4173 (2014).
10. M. Loncar and A. Faraon, *MRS Bull.* 38, 144 (2013).
11. J. Hagemeier, C. Bonato, T. Truong, H. Kim, G. Beirne, M. Bakker, M. van Exter, Y. Luo, P. Petroff, and D. Bouwmeester, *Opt. Express* 20, 24714-24726 (2012).
12. A. Faraon, C. Santori, Z. Huang, V. M. Acosta, and R. G. Beausoleil, *Phys. Rev. Lett.* 109, 033604 (2012).
13. T. Yoshie, A. Scherer, J. Hendrickson, G. Khitrova, H. M. Gibbs, G. Rupper, C. Ell, O. B. Shchekin, and D. G. Deppe, *Nature* 432, 200 (2004)
14. R. Schirhagl, K. Chang, M. Loretz & C.L. Degen, *Annu. Rev. Phys. Chem.* **65**, 83–105(2014).
15. B.-S. Song, S. Yamada, T. Asano, and S. Noda, *Opt. Express*, 19, 11084 (2011).
16. M. Radulaski, T. M. Babinec, S. Buckley, A. Rundquist, J Provine, K. Alassaad, G. Ferro, and J. Vučković, *Opt Express* 21, 26, 32623-32629 (2013).
17. N. T. Son, E. Sörman, W. M. Chen, M. Singh, C. Hallin, O. Kordina, B. Monemar, E. Janzén, and J. L. Lindström, *J. Appl. Phys.* 79, 3784 (1996).
18. V.Y. Bratus *et al.*, *Physica B* 404, 4739-4741 (2009).
19. A. F. Oskooi, D. Roundy, M. Ibanescu, P. Bermel, J. D. Joannopoulos, and S. G. Johnson, *Comput. Phys. Commun.* 181(3), 687–702 (2010).
20. S. Tomljenovic-Hanic, M. J. Steel, C. M. de Sterke, and J. Salzman, *Opt. Express* 14, 3556 (2006).
21. T. Akahane, T. Asano, B.-S. Song, S. Noda, *Nature* 425, 944 (2003)
22. L. C. Andreani, D. Gerace, and M. Agio, *Photonics Nanostr. Fundam. Appl.* 2, 103 (2004)
23. See supplementary material at **[URL will be inserted by AIP]** for details.
24. P. A. Khan, B. Roof, L. Zhou and I. Asesida, *J. Electron. Mater.* 30 212 (2001).
25. S.C. Ahn, S.Y. Hang, J.L. Lee, J.H. Moon, B.T. Lee, *Met. Mater. Int.* 10, 103 (2004).

26  M. McCutcheon, G. W. Rieger, I. W. Cheung, J. F. Young, D. Dalacu, S. Frédéric, P. J. Poole, G. C. Aers, and R. Williams, *Appl. Phys. Lett.* 87, 221110 (2005).
27  D. Englund, A. Faraon, I. Fushman, N. Stoltz, P. Petroff, and J. Vučković, *Nature* 450, 857 (2007).
28  M. Bosi, G. Attolini, M. Negri, C. Frigeri, E. Buffagni, C. Ferrari, T. Rimoldi, L. Cristofolini, L. Aversa, R. Tatti, R. Verucchi, *Journal of Crystal Growth* 383, 84-94 (2013).
29  J. Cardenas, M. Zhang, C. T. Phare, S. Y. Shah, C. B. Poitras, B. Guha, and M. Lipson, *Opt. Express* 21(14), 16882–16887 (2013).
30  T. Asano, B. S. Song, and S. Noda, *Opt. Express* 14, 1996 (2006).
31  M. Galli, S. L. Portalupi, M. Belotti, L. C. Andreani, L. O'Faolain, and T. F. Krauss, *Appl. Phys. Lett.* 94, 071101 (2009)..
32  Y. Tanaka, T. Asano, Y. Akahane, B. S. Song, and S. Noda, *Appl. Phys. Lett.* 82, 1661-1663 (2003).
33  L. Hiller, T. Stauden, R. M. Kemper, J. K. N. Lindner, D. J. As, and J. Pezoldt, *Proc. Mater. Sci.Forum* 717-720, 901 (2012).
34  R. Bose, D. Sridharan, G. S. Solomon, and E. Waks, *Appl. Phys. Lett.* 98, 121109 (2011).
35  K. Rivoire, Z. Lin, F. Hatami, W. T. Masselink, and J. Vuckovic, *Opt. Express* 17, 22609–22615 (2009).
36  P. M. Lundquist, H. C. Ong, W. P. Lin, R. P. H. Chang, J. B. Ketterson and G. K. Wong, *Appl. Phys. Lett.* 67, 2919 (1995).
37  S. Yamada, B.S. Song, S. Jeon, J. Upham, Y. Tanaka, T. Asano, and S. Noda, *Opt. Express* 39, 1768 (2014)
38  M. T. Rakher, L. Ma, O. Slattery, X. Tang, and K. Srinivasan, *Nat. Photonics* 4, 786–791 (2010).

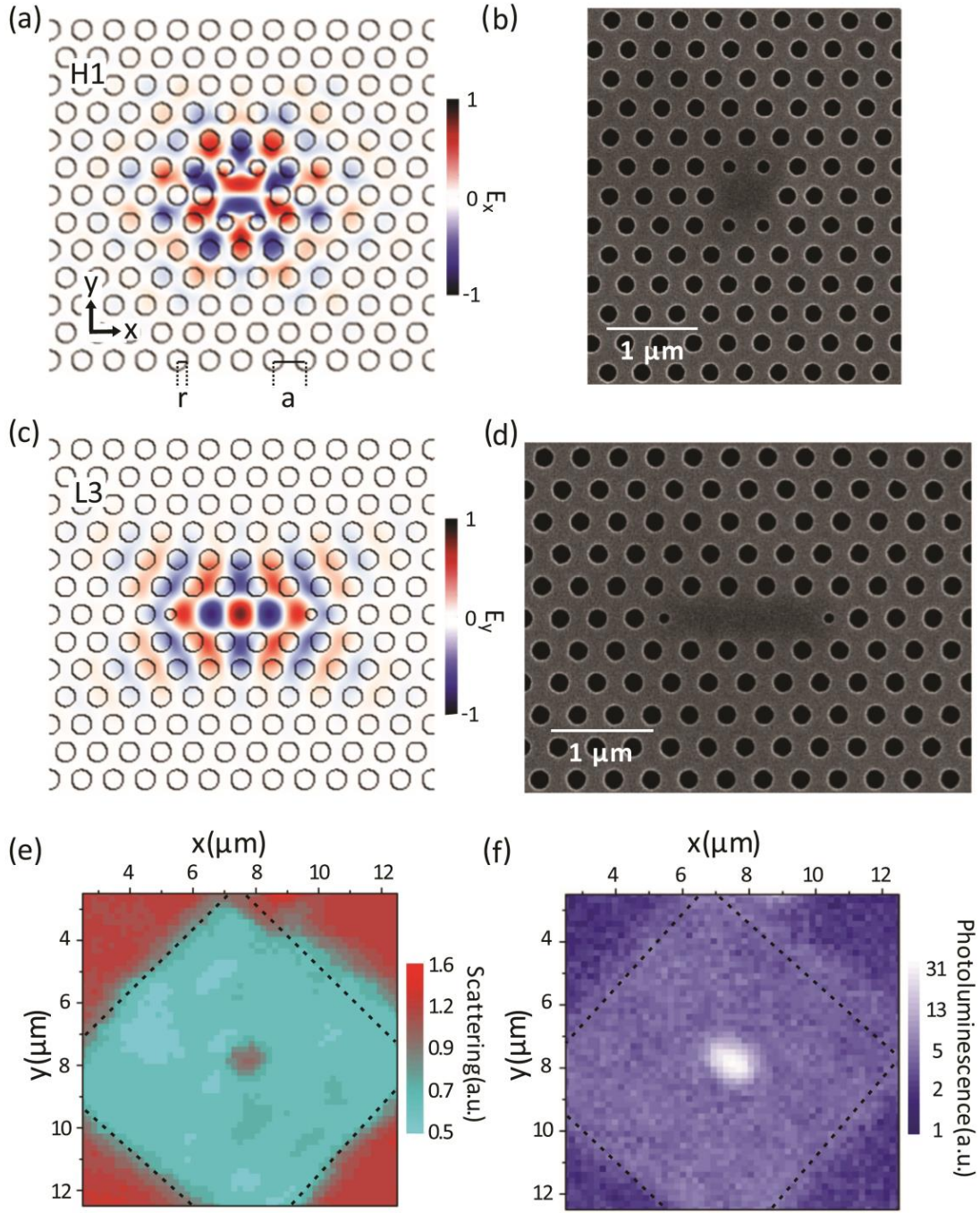

Fig 1: Photonic crystal cavities in 3C SiC. FDTD simulation of the electric field distribution $E_x$ (a) and SEM image (b) of the fabricated H1 cavity. The cavity Q is optimized by reducing the radius of the four nearest vertically adjacent holes and shifting the position of the horizontally adjacent holes. Electric field distribution $E_y$ (c) and SEM image (d) of the fabricated L3 cavity. The cavity Q is optimized by reducing the radius and shifting the position of the horizontally adjacent holes. Spatially resolved map of the off-resonant scattered laser light (e) and photoluminescence (f) from the L3 photonic crystal cavity mode at 20K.

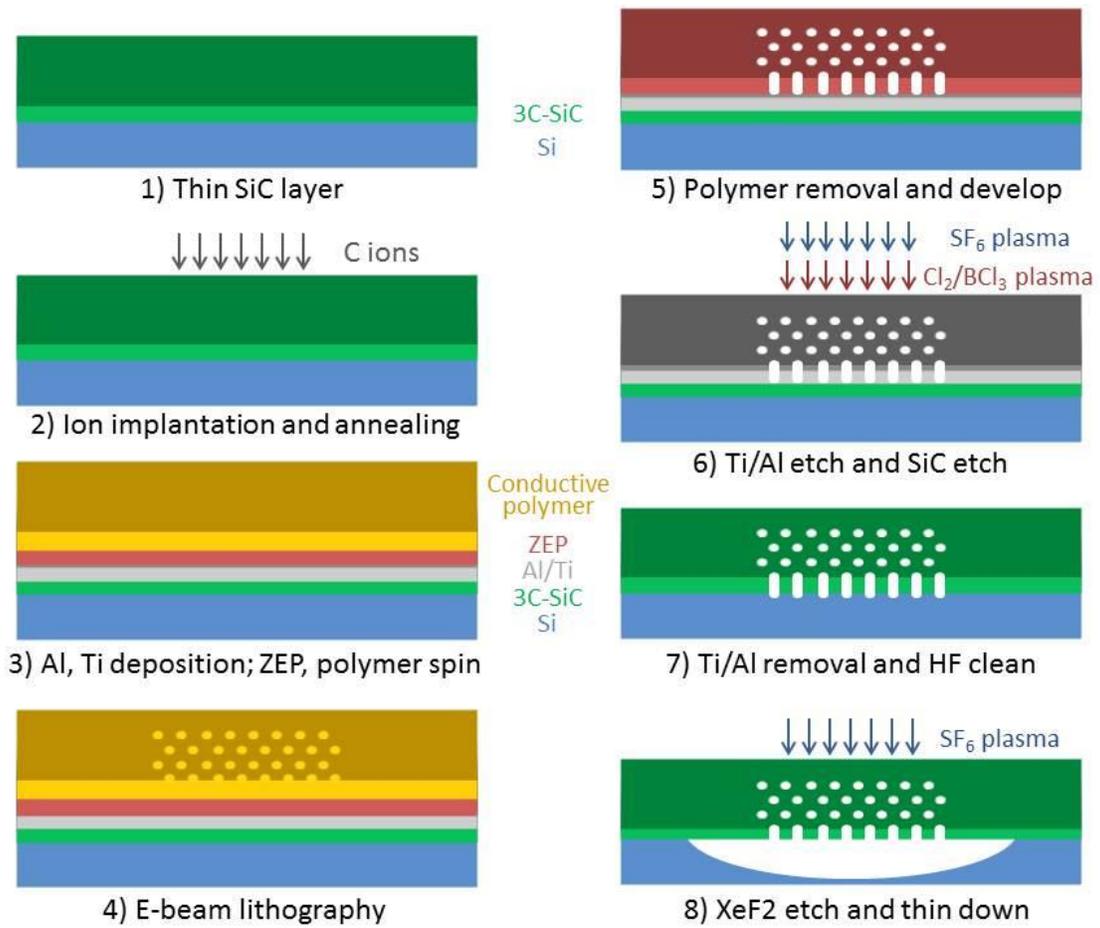

Fig 2: Photonic crystal fabrication process starting from a 300 nm thick 3C silicon carbide epilayer on a 100 mm silicon wafer.

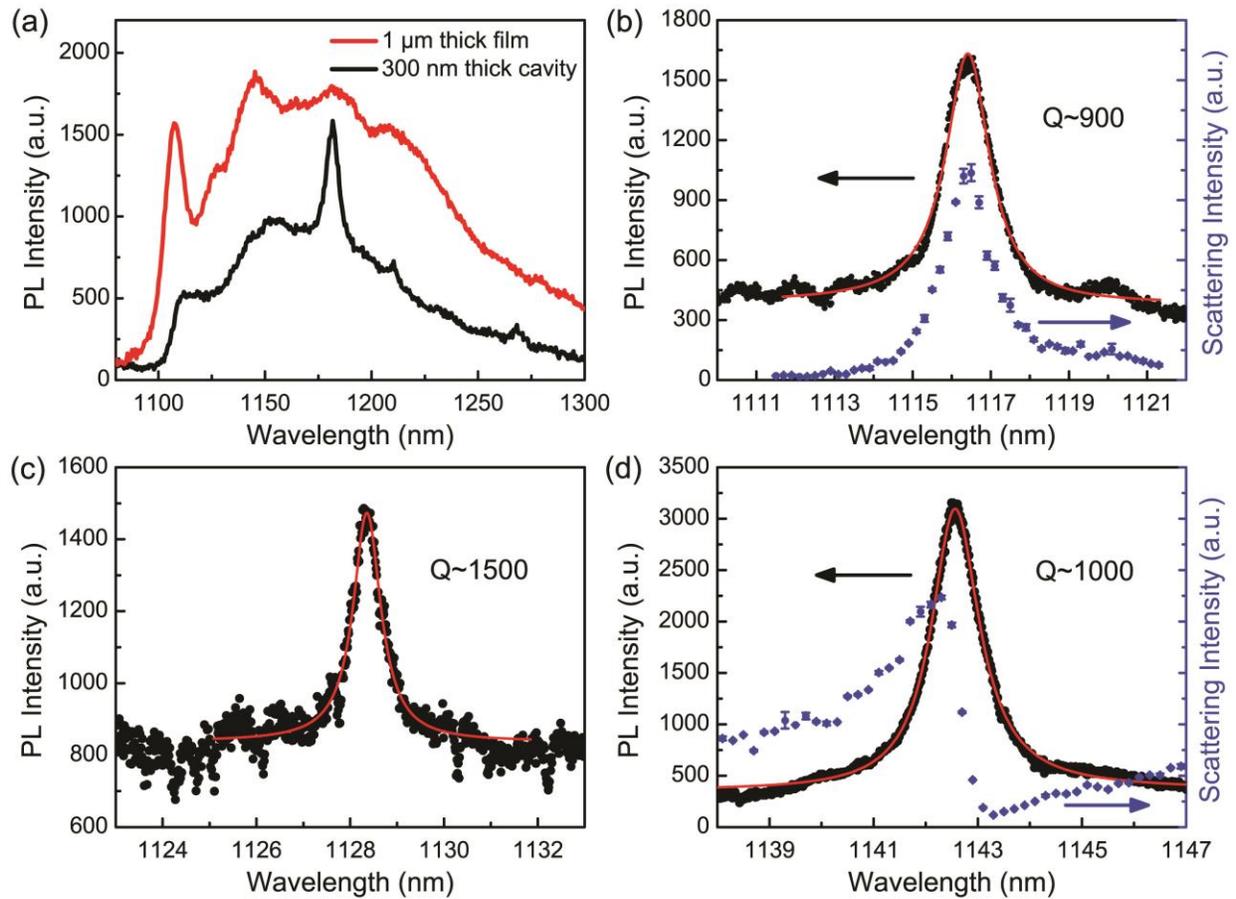

Fig 3 (a) Photoluminescence (PL) spectrum of Ky5 color centers in 3C SiC at 20K. The red curve shows the PL spectrum from an unprocessed 1 micron thick implanted and annealed 3C SiC slab, while the black curve shows the PL from an implanted and annealed 300 nm 3C SiC thin film after photonic crystal cavity patterning (cavity mode ~ 1180 nm). PL (black dots) and cross-polarized resonant scattering intensity (blue dots) measured for an (b) un-optimized L3 cavity, (c) optimized L3 cavity and (d) optimized H1 cavity. The red line on the PL spectrum represents a Fano-type best fit to extract the cavity Q factor.

TABLE I. Analysis of the cavity Q factor for various fabrication imperfections.

| Cavity Design | Ideal | Absorption[a] | Sidewall 88 deg | Sidewall 85 deg | Sidewall 82 deg | Hole Radius Variations | Hole Position Variations |
|---|---|---|---|---|---|---|---|
| L3 | 1,660 | 903 | 1,333 | 794 | 523 | 1,673 | 1,634 |
| L3 optimized | 17,084 | 9,153 | 7,363 | 2,198 | 1,059 | 16,383 | 16,090 |
| H1 optimized | 45,058 | 14,443 | 15,186 | 2,362 | 672 | 33,122 | 41,682 |

[a] Using the absorption coefficient from ref 29